\documentclass[pra, tw ocolumn, aps,floats,showpacs]{revtex4}

\usepackage{graphicx}
\usepackage{psfrag}

\begin{document}

\title{Probabilistic instantaneous quantum computation}

\author{{\v C}aslav Brukner$^1$, Jian-Wei Pan$^1$, Christoph Simon$^2$, Gregor Weihs$^1$, and Anton Zeilinger$^1$}
\affiliation{$^1$Institut f\"ur Experimentalphysik, Universit\"at Wien,  Boltzmanngasse 5, A--1090
Wien, Austria \\ $^2$Centre for Quantum Computation, University of
Oxford, Parks Road, Oxford OX1 3PU, United Kingdom}

\date{\today}

\begin{abstract}

The principle of teleportation can be used to perform a quantum
computation even before its quantum input is defined. The basic
idea is to perform the quantum computation at some earlier time
with qubits which are part of an entangled state. At a later time
a generalized Bell state measurement is performed jointly on the
then defined actual input qubits and the rest of the entangled
state. This projects the output state onto the correct one with a
certain exponentially small probability. The sufficient
conditions are found under which the scheme is of benefit.

\end{abstract}


\maketitle

Quantum computers \cite{nielsenchuangbook} -- a new type of machine that
exploits the quantum properties of information -- could perform
certain types of calculations with speedup over any
foreseeable classical computer. Quantum teleportation
\cite{bennett} -- one of the most basic information procedures in
quantum mechanics -- enables transmission and reconstruction of a
general quantum state over arbitrary distances. In 1997 Nielsen
and Chuang \cite{nielsenchuangarticle} have shown that the principle of
teleportation can be used to perform universal quantum computation
in a probabilistic fashion. Since then a substantial amount of
work has been done on generalizations and applications of the
idea. Gottesman and Chuang \cite{gottesman} developed
fault-tolerant constructions of quantum gates. Knill {\it et al.}
\cite{knill} apply the scheme to propose quantum computation with
linear optics together with single-photon detection and
single-photon sources. Further extensions have been developed
\cite{duer,vidal,huelga,nielsen,buzek}.

Here we consider the {\it time aspect} of
teleportation-based quantum computation. We show that with finite
probability the computational time of an arbitrary long
quantum computation can be saved completely. The basic idea is to use
quantum teleportation to perform the entire quantum computation
{\it even before} its quantum input is defined. This then allows
with certain exponentially small probability to obtain
the output of the computation immediately after its input is
given. The sufficient conditions are found under which the scheme is of
benefit.

Imagine that an engineer is given a certain problem of such a
complexity that in order to obtain its solution within reasonable
time she has to use a quantum computer. Suppose that the
conditions on the quantum computation are the following:
\begin{enumerate}
\item At time $t_1$ the engineer is given an input to the quantum
computation in an {\it arbitrary quantum state unknown} to
her.
\item The engineer is required to give the output of her
computation at time $t_2$. If however she is not sure that her
output is the right one (e.g. because her computer has not
finished the computation before $t_2$) {\it she is allowed not to
give any state}. Such a situation is denoted by "no answer".
\item The engineer is strongly advised not to over-estimate her
computational resources. By this we mean that the engineer's
choice to give no answer is not evaluated negatively, and that an
{\it incorrect result is evaluated more negatively than the
correct one is evaluated positively} (e.g. one may imagine that
she obtains P positive points for the correct result, 0 points for
no answer, and N negative points for an incorrect result, where N
is much larger than P.).
\end{enumerate}

\begin{figure}
\centering
\includegraphics[angle=0,width=8.9cm]{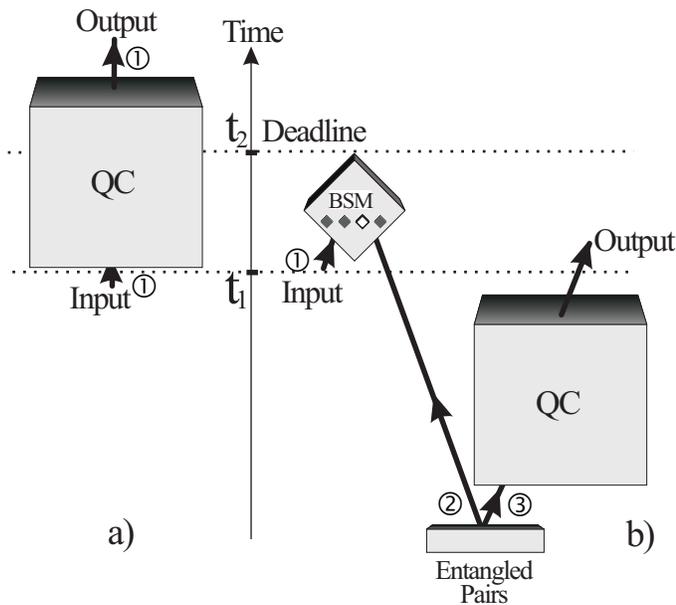} \caption{a)
Conventional scheme: At time $t_1$ the engineer is given the input
qubits 1 of the quantum computation (QC) in a quantum state
unknown to her. She feeds them into her quantum computer and
starts the computation. The computation is very time-consuming, so
that the quantum computer does not terminate before the deadline
at $t_2$. b) Scheme for instantaneous quantum computation: At a
time earlier than $t_1$ the engineer has fed qubits 3, which are
each maximally entangled with one qubit 2, into her quantum
computer and has done the computation. At the later time $t_1$
when the input qubits 1 are given to her the engineer performs a
Bell-state measurement (BSM) on each pair of qubits 1 and 2 and
projects qubits 3 onto a corresponding state. In a certain exponentially small
fraction of cases the computational time is saved completely as she
immediately knows that qubits 3 are projected onto the output
state resulting from the correct input one.} \vspace{-0.2cm}
\label{exam}
\end{figure}

For the purpose of evaluating the engineer's computation, one may
imagine that whenever the engineer decides to give the output of
her computation it is subjected to a kind of
"check-measurement" in a basis in which one of the basis states is
the right output state. Denote the outcome corresponding to the
right output by $O$. Then, only if the measurement gives the
result $O$ does the engineer gain P points, otherwise she loses N
points.

What are useful typical situations where the input of the
computation  is an arbitrarily and unknown quantum state as required in 1?
All known quantum computation algorithms start with
classical inputs somewhere along the line. To our knowledge only
the quantum simulations \cite{lloyd} and certain quantum
cryptography protocols start with quantum inputs. Also
realistic situations where the duration of a quantum
computation is restricted by a deadline are not typical.
Is there any example for the situations described by points 1.-3?
We give an explicit example for such situation for quantum simulations
later in the text. Here we note that the input in universal quantum operations,
like universal quantum cloning \cite{buzekclone} or universal quantum universal-NOT operation \cite{buzekNOT}
is an arbitrary and unknown quantum state \cite{prva}.

We now consider possible strategies which the engineer can follow.
Normally, after the engineer gets the input $n$ qubits (qubits 1)
for the quantum computation, she feeds them into her computer and
starts the quantum computation. Now assume the quantum computation
is very time-consuming, so that this procedure is not fast enough
and the quantum computer does not terminate before the deadline at
$t_2$, as illustrated in Fig. \ref{exam}a. In such a situation the
engineer, for instance, can decide not to give an answer, which
results in a total of zero points. Alternatively, she can choose
any state at random and give this as the output of her
computation. This however leads to a high negative score because
the probability of $1-(1/2)^n$ not to obtain the result $O$ in the
check-measurement is higher than the one of $1/2^n$ to obtain this
result for a n-qubit state $(n>1)$ chosen at random.

We will now show that there is an alternative strategy where the
engineer can have a positive score. There she can obtain the
{\it exact} output state of an arbitrarily
long quantum computation instantaneously with some probability.
Most importantly, she knows when she obtains the correct output state and thus
can pass it on for evaluation. The strategy uses teleportation-based
quantum computation \cite{nielsenchuangarticle}.

Quantum teleportation \cite{bennett} is the
transmission and reconstruction over arbitrary distances of the
state of a quantum system. During teleportation, an initial system
in the state that is to be transferred and one of a pair of
entangled subsystems are subjected to a Bell state measurement,
such that the second subsystem of the entangled pair acquires the
state of the initial system. The later subsystem is brought into
the state of initial system by an accordingly chosen
transformation after receiving via classical communication channel
the information which of the Bell-state results was obtained.

Now imagine that the engineer has $n$ entangled pairs of qubits 2
and 3. She can feed members of the entangled pairs (qubits 3) into
her quantum computer long before $t_1$, when the actual input for
her computation is given to her, as illustrated in Fig.
\ref{exam}b. This means that during the computation the qubits 3
in her quantum computer are entangled to the qubits 2. At some
point her computation will terminate and output qubits 3.

As soon as she obtains the input qubits 1 the engineer performs
the Bell state measurement on each of $n$ pairs of qubits 1 and 2.
In $(1/4)^n$ cases the whole state of qubits 3 is projected onto
the state resulting from the correct input and she does not have
to perform any additional transformation on qubits 3. In the
remaining $1-(1/4)^n$ cases, the result of the engineer's Bell
state analysis will not be the right one. However, in a situation
(as defined by conditions 1-3 given above) where it is of
advantage to both have the correct output state for a problem and to
know it at a very early time even if only with small probability,
our scheme is of benefit.

We now consider also the remaining cases. In the usual
teleportation procedure, the engineer would have to perform single-qubit
unitary transformation on her qubits 3. But now she has already
fed them into her quantum computer. Because the set of unitary
transformations necessary to complete the usual teleportation
procedure and her computation do not always commute, in general
she has to invert the full quantum computation performed so far,
perform the single-qubit transformations required by the teleportation
procedure, and start the quantum computation again \cite{footnote}.
The scheme is thus only efficient for the case where no transformation are done after the Bell-state
measurement. This results
in a total computational time twice as long as the time needed for
the conventional computation (if we neglect the time required for
the single-qubit transformations). This means that in
a fraction $1-(1/4)^n$ of all cases in general our scheme will not
help the engineer to meet her deadline. In those cases the best
strategy for her is to give no answer.

But in the succesful case, depending on the length of her computation, the scheme for
instantaneous computation may constitute an enormous gain in time,
which might be decisive in situations as specified by conditions 1-3 listed above.
Following the evaluation criterion 3, the averaged point score gained in our
scheme is $S_{inst}=P (1/4)^n$, which exceeds both the score of
$S_{no}=0$ if the engineer constantly provides no answer, and the
score of $S_{rand}=P (1/2^n)- N (1-1/2^n)$, if she constantly
chooses the output at random.

It is important to note that in our protocol the resources are
consumed even in the cases when the engineer does not provide an
answer. The reason is that the engineer's computation has to be
run each time thus using up $n$ entangled pairs. Moreover, since
the probability of success is $(1/4)^n$, there is an expected
exponential cost of entangled pairs to get a single successful run
of the computation. If, for example, each computation requires one
hour, but $n_0=1000$ attempts must be made (which is the
approximate number $4^n$ for $n=5$), the company that distributes
entangled pairs has to be paid for 1000 hours, not just one hour.
There are however situations where it clearly still pays off for
the engineer to follow our scheme. These are when the engineer's
gain P (which in our example could correspond to a certain amount
of money) is larger than the overall costs incurred over the
average number of attempts required to get one successful run,
i.e. when $P > n_0 C$, where $C$ are the costs of a single run.

It should be stressed again that the protocol based on
teleportation is only reasonable to consider if the engineer's
input is in an arbitrary quantum state unknown to her. In
contrast, if the input is a {\it classical} one, then a simple protocol
is possible which even do not require entanglement but has a
success rate higher than the teleportation-based one. Suppose that
the input still is in the state unknown to the engineer, but this
state is one of $2^n$ (orthogonal) states which are known in
advance to the engineer. Then the engineer simply can perform the
computation in advance choosing any of the $2^n$ states as the
input. Later, when the actual input is given to her, she can
measure it in the basis build out of the $2^n$ states. On the
basis of the result obtained she then knows with certainty whether
she had performed the computation with the correct input or not.
With probability $1/2^n$ her input will be the correct one, when
she passes on her output; otherwise she provides no answer.

Another situation where requirement 1 is not fulfilled is when the
input is in an arbitrary quantum state, but this state becomes
{\it known} to the engineer at the time the input is given to her.
In such a situation a protocol is possible which does require
entanglement but not Bell-state measurement and has a success rate
higher than the teleportation-based one. The protocol is based on
the ''remote state preparation" \cite{xxx}. The engineer starts
with $n$ pairs of entangled qubits 2 and 3 and processes the
computation on qubits 3 in advance. When she learns what is the
actual input state, she performs such a measurement on qubits 2 to
project qubits 3 onto the correct output state. The probability of success
for the scheme is $1/2^n$.

There are also alternative schemes \cite{alternative} which
fulfill requirement 1 but give an output state only very close to
the correct one and, thus, still with a finite probability to show
the wrong result when measured in the check-measurement. In
contrast with our scheme, there the engineer cannot infer with
certainty whether her output state is the correct one or not.
Therefore if she decides to pass on her output state, there will
always be a certain probability not to obtain the result $O$ in
the check-measurement, which consequently leads to a negative
average point score. Clearly such a scheme can never be better
than ours, for penalty N for a wrong result sufficiently larger
than gain P for a correct one.

The teleportation-based protocol scheme can also be applied in
cases where, for some reason, parts of a quantum computation are
performed at distant locations \cite{huelga,anton}. Imagine two
people, Alice and Bob, in two distant locations, each of them
performing part of a common quantum computation under conditions
1-3. Suppose that the output qubits (identified with qubits 1 in
our scheme) of Alice's quantum computer are an essential input for
Bob's computation. Suppose also that Bob's part of computation is
very time-consuming.

Imagine that entangled pairs of qubits (identified with qubits 2
and 3 in our scheme) are distributed to Alice and Bob over some
quantum network in advance. Bob now can immediately feed his
members of the entangled pairs (qubits 3) into his quantum
computer. Thus, he can start his time-consuming computation long
before the input of Alice's part of the computation is given at
$t_1$. At some point after $t_1$ Alice's computation will
terminate and output the qubits 1 that Bob needs. They can now
proceed as in the usual quantum teleportation procedure. Alice
performs the Bell state measurement on each of the pairs qubits 1
and 2. Because Bob has been able to start his computation much
earlier than Alice, it is natural to assume that Bob's computation
has terminated by the time Alice's call reaches him. Note that
even if Alice had not done her computation before Bob did, or even
started, his computation, if separation between them is
large enough, he has enough time to do his calculation
before Alice's call arrives. They proceed now analogous to the
scheme given in the text above. If Alice tells him that the result
of her Bell state measurement was the good one, he immediately
knows that the output of his quantum computer is the correct one.
Only in those cases he passes on the output of his computation,
otherwise he provides no answer. This again results in averaged
point score of $S_{inst}$ as given above.

We now give an explicit example for a situation described by conditions 1.-3.
Although it admittedly might appear to be contrived to some extent, we are
convienced that the existence of even just one example points at novel conceptual
possibilities. The example is based on quantum simulations \cite{lloyd}.
Imagine that our engineer is still a student of engineering.
Suppose that she has a series of exams which each consists of
performing a specific quantum simulation. Which ones is known to
her in advance (e.g. the particular ''exam question" is given to students a
day before the exam takes place). Further, suppose that the
process of the examination and the evaluation of its success is
executed according to the rules 1-3. The reason
that the input is in an arbitrary quantum state unknown to students is
to prevent them to find out the correct solution just on the basis
of the known solutions from the previous generations of students.
Now assume that in order to pass the exams students need to
have a positive score of points averaged over the series. The
points are given on the basis of the check-measurement performed
by an examiner who knows the correct output state and thus can choose the
adequate measurement basis. In order to ascertain the best of them the
examiner faces the students with seemingly unsolvable problem of
performing a series of quantum simulations which each consume more time than the
duration of a single exam. Obviously only if our student applies the
teleportation-based scheme she can expect to have the correct result in a
certain fraction of the total number of the exams and thus having a positive score of points
to pass the exams.

In summary, we have shown that, using the principle of
teleportation, one may achieve instantaneous quantum computation
where: (a) one can obtain the output of an arbitrarily long
computation with non-zero probability in zero computational time,
i.e. immediately after its quantum input is defined, and (b) one
knows when the output is correct without knowing the output. We identify the sufficient
conditions under which the scheme is more efficient than any alternative one.
We suggest that the ability to perform a quantum computation even
before its input is defined derives from the possibility to
process information represented by a quantum state without need of
reading the state beforehand.

We would like to thank Terry Rudolph and Tomas Tyc for helpful
comments and discussions. This work has been supported by the
Austrian Science Foundation FWF, Project No. F1506 and by the QIPC
Program of the European Union.



\begin{thebibliography}{99}
\bibitem{prva} Note that situation of an eavesdropper in certain quantum cryptography
protocols might be similar to the situation of our engineer. For example, the input for eavesdropper operations is
usually a quantum state unknown to her. Also one can assume that her
operations need to be performed in a restricted time, for example, for the reason not to be detected by the
parties following the protocol by noticing a delay in the exchange
of the quantum states between the parties, which could be caused by the eavesdropper's
action. Of course this has nothing to do with
security based on detecting the state disturbance produced by an eavesdropping strategy.
\bibitem{alternative} We give an example for the alternative scheme
for the case of a one-qubit computation. At a time earlier than
$t_1$ the engineer has taken any state, say state $|0\rangle$, as
an input of her computation and has performed the computation on
it. At the later time $t_1$ when the actual input is given to her,
she measures it in the basis $\{|0\rangle,|1\rangle\}$. If the
result is $|0\rangle$, she passes on the output state of her computation and if
the result is $|1\rangle$, she passes on the state which is obtained by
applying the optimal universal NOT \cite{buzekNOT} on the output. The
engineer uses the optimal universal NOT because she does not know
the output state of her computation.
\end{thebibliography}
\end{document}